\documentclass[12pt]{article}
\usepackage{html}
\usepackage{epsf}
\usepackage[pdftex]{graphicx}
\usepackage{graphicx}
\usepackage{amssymb}
\usepackage{amsmath}
\usepackage{hyperref}
\begin{document}
\begin{titlepage}
\includegraphics[width=150mm]{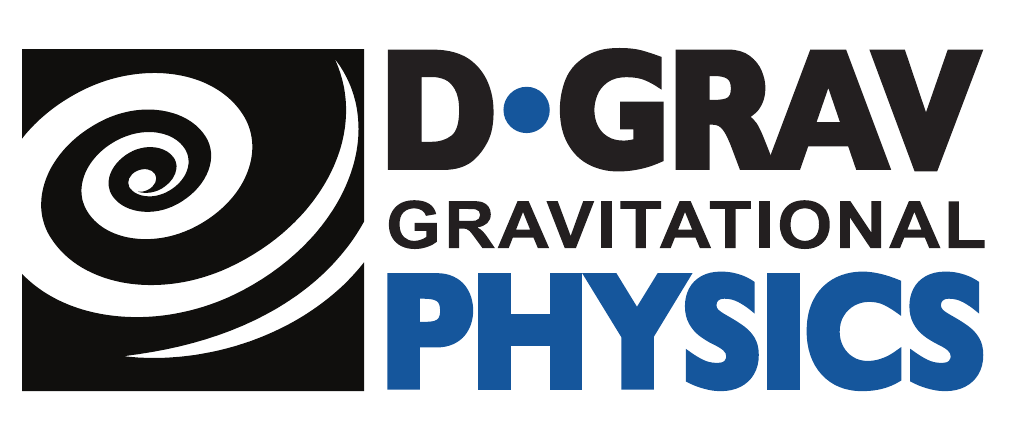}
\begin{center}
{ \Large {\bf MATTERS OF GRAVITY}}\\ 
\bigskip
\hrule
\medskip
{The newsletter of the Division of Gravitational Physics of the American Physical 
Society}\\
\medskip
{\bf Number 50 \hfill December 2017}
\end{center}
\begin{flushleft}
\tableofcontents
\end{flushleft}
\end{titlepage}
\vfill\eject
\begin{flushleft}
\section*{\noindent  Editor\hfill}
David Garfinkle\\
\smallskip
Department of Physics
Oakland University
Rochester, MI 48309\\
Phone: (248) 370-3411\\
Internet: 
\htmladdnormallink{\protect {\tt{garfinkl-at-oakland.edu}}}
{mailto:garfinkl@oakland.edu}\\
WWW: \htmladdnormallink
{\protect {\tt{http://www.oakland.edu/physics/Faculty/david-garfinkle}}}
{http://www.oakland.edu/physics/Faculty/david-garfinkle}\\

\section*{\noindent  Associate Editor\hfill}
Greg Comer\\
\smallskip
Department of Physics and Center for Fluids at All Scales,\\
St. Louis University,
St. Louis, MO 63103\\
Phone: (314) 977-8432\\
Internet:
\htmladdnormallink{\protect {\tt{comergl-at-slu.edu}}}
{mailto:comergl@slu.edu}\\
WWW: \htmladdnormallink{\protect {\tt{http://www.slu.edu/arts-and-sciences/physics/faculty/comer-greg.php}}}
{http://www.slu.edu/arts-and-sciences/physics/faculty/comer-greg.php}\\
\bigskip
\hfill ISSN: 1527-3431


\bigskip

DISCLAIMER: The opinions expressed in the articles of this newsletter represent
the views of the authors and are not necessarily the views of APS.
The articles in this newsletter are not peer reviewed.

\begin{rawhtml}
<P>
<BR><HR><P>
\end{rawhtml}
\end{flushleft}
\pagebreak
\section*{Editorial}

The next newsletter is due June 2018.  Issues {\bf 28-50} are available on the web at
\htmladdnormallink 
{\protect {\tt {https://files.oakland.edu/users/garfinkl/web/mog/}}}
{https://files.oakland.edu/users/garfinkl/web/mog/} 
All issues before number {\bf 28} are available at
\htmladdnormallink {\protect {\tt {http://www.phys.lsu.edu/mog}}}
{http://www.phys.lsu.edu/mog}

Any ideas for topics
that should be covered by the newsletter should be emailed to me, or 
Greg Comer, or
the relevant correspondent.  Any comments/questions/complaints
about the newsletter should be emailed to me.

A hardcopy of the newsletter is distributed free of charge to the
members of the APS Division of Gravitational Physics upon request (the
default distribution form is via the web) to the secretary of the
Division.  It is considered a lack of etiquette to ask me to mail
you hard copies of the newsletter unless you have exhausted all your
resources to get your copy otherwise.

\hfill David Garfinkle 

\bigbreak

\vspace{-0.8cm}
\parskip=0pt
\section*{Correspondents of Matters of Gravity}
\begin{itemize}
\setlength{\itemsep}{-5pt}
\setlength{\parsep}{0pt}
\item Daniel Holz: Relativistic Astrophysics,
\item Bei-Lok Hu: Quantum Cosmology and Related Topics
\item Veronika Hubeny: String Theory
\item Pedro Marronetti: News from NSF
\item Luis Lehner: Numerical Relativity
\item Jim Isenberg: Mathematical Relativity
\item Katherine Freese: Cosmology
\item Lee Smolin: Quantum Gravity
\item Cliff Will: Confrontation of Theory with Experiment
\item Peter Bender: Space Experiments
\item Jens Gundlach: Laboratory Experiments
\item Warren Johnson: Resonant Mass Gravitational Wave Detectors
\item David Shoemaker: LIGO 
\item Stan Whitcomb: Gravitational Wave detection
\item Peter Saulson and Jorge Pullin: former editors, correspondents at large.
\end{itemize}
\section*{Division of Gravitational Physics (DGRAV) Authorities}
Chair: Peter Shawhan; Chair-Elect: 
Emanuele Berti; Vice-Chair: Gary Horowitz. 
Secretary-Treasurer: Geoffrey Lovelace; Past Chair:  Laura Cadonati; Councilor: Beverly Berger
Members-at-large:
Duncan Brown, Michele Vallisneri, Kelly Holley-Bockelmann, Leo Stein, Lisa Barsotti, Theodore Jacobson.
Student Members: Megan Jones, Cody Messick.
\parskip=10pt

\vfill\eject

\section*{\centerline
{we hear that \dots}}
\addtocontents{toc}{\protect\medskip}
\addtocontents{toc}{\bf DGRAV News:}
\addcontentsline{toc}{subsubsection}{
\it we hear that \dots , by David Garfinkle}
\parskip=3pt
\begin{center}
David Garfinkle, Oakland University
\htmladdnormallink{garfinkl-at-oakland.edu}
{mailto:garfinkl@oakland.edu}
\end{center}

{\bf Kip Thorne, Rainer Weiss, and Barry Barish have been awarded the 2017 Nobel Prize in Physics}

Kip Thorne, Rainer Weiss, and Barry Barish have been awarded the 2017 Fudan-Zhongzhi Science Award.

Dennis Coyne, Peter Fritschel, and David Shoemaker have been awarded the Berkeley Prize of the American Astronomical Society.

Alessandra Buonanno has been awarded the Leibniz Prize.

Alain Brillet and Thibault Damour have been awarded the CNRS Gold Medal.

Charles Bennett, Gary Hinshaw, Norman Jarosik, Lyman Page, David Spergel, and the WMAP Science Team have been awarded the 2018 Breakthrough Prize in Fundamental Physics.

Eugene Parker has been awarded the APS Medal for Exceptional Achievement in Research.

The first multimessenger detection of a neutron star merger was Physics World's breakthrough of the year for 2017.

Gabrielle Allen, John Baker, Jolien Creighton, Charles Evans, Eric Gustafson, Daniel Holz, Vuk Mandic, and Yasunori Nomura have been elected APS Fellows.

Hearty Congratulations!

\vfill\eject
\section*{\centerline
{April APS meeting}
}
\addtocontents{toc}{\protect\medskip}
\addcontentsline{toc}{subsubsection}{
\it APS April Meeting, by David Garfinkle}
\parskip=3pt
\begin{center}
David Garfinkle, Oakland University
\htmladdnormallink{garfinkl-at-oakland.edu}
{mailto:garfinkl@oakland.edu}
\end{center}

We have a very exciting DGRAV related program at the upcoming APS meeting April 14-17 in Columbus, Ohio.  Our Chair-elect, Emanuele Berti, did an excellent job of putting together this program.
\vskip0.25truein
DGRAV related plenary talks include the following:\\

There will be a plenary session with talks by Nobel Prize recipients Weiss and Barish and APS Medal winner Eugene Parker.  A plenary talk by Marcelle Soares-Santos will summarize the electromagnetic signatures of GW170817.  Another plenary talk, by Anne Archibald, will report bounds on the equivalence principle from observations of PSR J0337+1715 (a millisecond pulsar in a hierarchical triple system with two other stars).
\vskip0.25truein
The DGRAV sponsored invited sessions are\\

{\bf RECENT LIGO/Virgo RESULTS} (Chair: Laura Cadonati)\\
Jenne Driggers, {\it Advanced LIGO and Advanced Virgo}\\
Jolien Creighton, {\it Gravitational Waves from Compact Binary Coalescences Detected during the Second LIGO/Virgo Observing Run}\\
Sinead Walsh, {\it Searches for continuous gravitational waves in advanced LIGO}\\

{\bf DEVELOPMENTS IN GRAVITATIONAL THEORY} (Chair: David Garfinkle)\\
Beatrice Bonga, {\it Surprising consequences of a positive cosmological constant}\\
Chad Galley, {\it Gravitational Waves from Compact Binaries in an Effective Field Theory Approach}\\
Robert Wald, {\it Black Holes Cannot be Over-Charged or Over-Spun}\\

{\bf QUANTUM ASPECTS OF GRAVITATION} (Chair: Gary Horowitz)\\
Wolfgang Wieland, {\it Loop Quantum Gravity and the Quantisation of Null Surfaces}\\
Aron Wall, {\it How Low Can the Energy Density Go?}\\
Mark van Raamsdonk, {\it Classical gravitational physics from quantum information theory}\\

{\bf THIRD GENERATION GW DETECTORS } (Chair: Peter Shawhan)\\
Matt Evans, {\it  Third-generation gravitational-wave detectors - how we will reach the edge of the Universe}\\
Salvatore Vitale, {\it The Scientific Potential of Third Generation Gravitational Wave Detectors}\\
Ho Jung Paik, {\it SOGRO: Superconducting Tensor Detector for Mid-Frequency Gravitational Waves}\\ 

{\bf UNVEILING MASSIVE BLACK HOLES } (Chair: Kelly Holley-Bockelmann)\\
Priyamvada Natarajan, {\it Unveiling the first black holes}\\
Enrico Barausse, {\it The quest for low frequency gravitational waves}\\
Cole Miller, {\it Arguments For and Against Intermediate-Mass Black Holes}\\

{\bf GW SOURCES – COMPACT BINARY FORMATION SCENARIOS} (Chair: Daniel Holz)\\
(co-sponsored with DAP)\\
Michela Mapelli, {\it Dynamics versus Isolated Binary Evolution: Place Your Bets}\\
Davide Gerosa, {\it What do LIGO's Black Holes Remember?}\\
Will Farr, {\it Uncovering the Formation of Compact Objects Through Gravitational Wave Observations}\\

{\bf LOW-FREQUENCY GRAVITATIONAL WAVE ASTRONOMY} (Chair: Michele Vallisneri)\\ 
(co-sponsored with DAP)\\
Ira Thorpe, {\it Observing the milliHertz Gravitational Wave Sky with LISA}\\
Steve Taylor, {\it Pulsar Timing Arrays: New Advances Toward Detecting Low-frequency Gravitational Waves}\\
Maura McLaughlin, {\it Pulsar Timing Arrays: Building a Low-Frequency Gravitational Wave Detector}\\

{\bf THEORY OF EM COUNTERPARTS TO GW EVENTS}  (Chair: Cole Miller)\\
(co-sponsored with DAP)\\
Davide Lazzati, {\it Cocoons, Structured Jets, and the non-Thermal Emission of Binary Neutron Star Mergers}\\
Brian Metzger, {\it Kilonovae from Binary Neutron Star Mergers}\\
Enrico Ramirez-Ruiz, {\it Electromagnetic Transients from Compact Binary Mergers}\\

{\bf FIREBALL EMISSION FROM NS-NS MERGERS}  (Chair: Judith Racusin)\\
(co-sponsored with DAP)\\
Anthony Piro, {\it Shedding Light on Gravitational Waves}\\
Eric Burns, {\it GW170817 and GRB 170817A}\\
Gregg Hallinen, {\it The Radio Afterglow of the Neutron Star Merger GW170817}\\

{\bf GWs \& DM SEARCHES} (Chair: Marc Kamionkowski)\\
(co-sponsored with DAP and DPF)\\
Lam Hui, {\it Ultra-light Axion Dark Matter}\\
Masha Baryakhtar, {\it Searching for Ultralight Particles with Black Holes and Gravitational Waves}\\
Ilias Cholis, {\it Primordial Black Holes as Gravitational Wave Sources}\\

{\bf BINARY MERGER SIMULATIONS} (Chair: Frans Pretorius)\\
(co-sponsored with DAP and DCOMP)\\
Luis Lehner, {\it New Vistas in Binary Black Holes}\\
Vasileios Paschalidis, {\it Binary Neutron Star Mergers: Nature's Lab for Astrophysics and Nuclear Physics}
\\
Francois Foucart, {\it Modeling Kilonovae Using Neutron Star Merger Simulations : Current Status and Uncertainties}\\

{\bf TESTS OF GENERAL RELATIVITY } (Chair: Clifford Will)\\
(co-sponsored with DAP and GPMFC)\\
Jason Hogan, {\it Atom Interferometry for Equivalence Principle Tests and Gravitational Wave Detection}\\
Brian D'Urso, {\it Levitated Optomechanics for Precision Gravitational Measurements}\\
Manuel Rodrigues, {\it The First Scientific Results of MICROSCOPE, the Space Test of the Equivalence Principle}\\

{\bf THE CHAPEL HILL CONFERENCE}  (Chair: Daniel Kennefick)\\
(co-sponsored with FHP)\\
Peter Saulson, {\it How Felix Pirani Launched the Effort to Detect Gravitational Waves}\\
Dean Rickles, {\it Behind the Scenes of the Chapel Hill Conference}\\
Dieter Brill, {\it Wheeler and I}\\
Joshua Goldberg, {\it US Airforce Support of General Relativity: Chapel Hill 1957 and Beyond}\\

\vskip0.25truein
The Focus Sessions sponsored by DGRAV (co-sponsored with DAP) are\\
{\bf
Optical Counter-parts to Gravitational Wave Events\\
NICER\\
}

\vfill\eject
\section*{\centerline
{Town Hall Meeting}
}
\addtocontents{toc}{\protect\medskip}
\addcontentsline{toc}{subsubsection}{
\it Town Hall Meeting, by Emanuele Berti}
\parskip=3pt
\begin{center}
Emanuele Berti, University of Mississippi
\htmladdnormallink{eberti-at-olemiss.edu}
{mailto:eberti@olemiss.edu}
\end{center}

Gravitational wave astronomy is finally a reality. As a community, we should strive to make the best of the wealth of data that will be collected in the next few years. The first multi-messenger observation of a binary neutron star merger (GW170817) was one of the biggest science stories of 2017, and binary black hole detections are becoming routine. However there is still a lot of work needed to coordinate the efforts of instrumentalists, data analysts, astrophysicists and gravitational theorists.

For this reason, Beverly Berger and Manuela Campanelli proposed to organize a Town Hall Meeting on ``Gravitational Wave Theory and Simulations in the Era of Detections'' at the upcoming APS April Meeting in Columbus, OH. The Town Hall Meeting will be sponsored by DGRAV and co-sponsored by DAP.

The idea is to begin a conversation to identify the theory and simulations most needed to exploit expected gravitational wave detections by LIGO and Virgo in the next two science runs, O3 and O4. Possible discussion topics include the end state of binary neutron star (or neutron star/black hole) mergers, tests of general relativity, black hole spectroscopy, compact object spins, extracting information about the neutron star equation of state, and so on.

The LIGO Scientific Collaboration will present the anticipated evolution of noise power spectral densities in the next few years, together with the signals that were already detected. We will discuss implications of improvements to theory and simulations, as well as the impact that sensitivity improvements will have on theory and simulations (what might be ruled out or in? what new research topics should be prioritized?)

The goal is to identify a reasonably complete list of theory and simulation research topics that might inform (and be informed by) gravitational wave events detected in O3 and O4.

We would like to call for contributed papers from the LSC, theory, and simulation communities on research that might be relevant to this Town Hall Meeting, providing the opportunity for a short (1-3 minute!) summary slide for each presenter at the Town Hall Meeting. The Town Hall Meeting is expected to take place after the DGRAV business meeting. Please email Emanuele Berti (\htmladdnormallink{\protect {\tt{eberti-at-olemiss.edu}}}
{mailto:eberti@olemiss.edu}) and Beverly Berger (\htmladdnormallink{\protect {\tt{beverlyberger-at-me.com}}}
{mailto:beverlyberger@me.com}) if you have submitted, or plan to submit, a Town Hall-related abstract.



\vfill\eject

\section*{\centerline
{﻿Gravity and Black Holes Conference and Symposium}}
\addtocontents{toc}{\protect\medskip}
\addtocontents{toc}{\bf Conference Reports:}
\addcontentsline{toc}{subsubsection}{
\it  Hawking Conference, by Harvey Reall and Paul Shellard}
\parskip=3pt
\begin{center}
Harvey Reall and Paul Shellard, Cambridge University 
\htmladdnormallink{hsr1000-at-cam.ac.uk}
{mailto:hsr1000@cam.ac.uk}
\htmladdnormallink{E.P.S.Shellard-at-damtp.cam.ac.uk}
{mailto:E.P.S.Shellard@damtp.cam.ac.uk}
\end{center}

This International Conference and associated Public Symposium, organized by the Centre for Theoretical Cosmology (DAMTP, Cambridge), was held in honour of the 75th birthday of Stephen Hawking. About 180 participants attended the scientific conference from 3-5 July, while the preceding symposium, on Sunday July 2 was attended by 450 people and reached a much larger online audience via a live webcast.

The Public Symposium consisted of four popular lectures delivered by well-known science communicators. Brian Cox eloquently described the development of modern concepts in physics which have defined the perception of our place in the Universe.   Gabriela Gonz\'alez described the exciting discovery of gravitational waves from merging black holes by the LIGO team, and described the prospects for a new era of gravitational wave astronomy.   Martin Rees took the audience on a fascinating journey from small-scale exoplanets out beyond the observed universe into the multiverse.   Finally, Stephen Hawking recounted his own life in physics and his contributions to momentous developments in our understanding of black holes and cosmology, which continue to remain at the heart of key theoretical and experimental programmes to the present day. 

The scientific conference which followed was roughly divided across days into topical themes. The talks on Monday were mostly devoted to cosmology. Both Eiichiro Komatsu and Hiranya Peiris explained how observational data from the CMB and galaxy surveys can be used to constrain theories of inflation and fundamental physics, while also describing future prospects for experimental programs seeking to discover primordial gravitational waves (tensor modes) in CMB polarization data. Slava Mukhanov, argued that standard inflation makes definite predictions, in contrast to some criticisms of the theory, and that current data does not necessitate elaborate extensions to inflation. After a brief overview, Andrei Linde discussed “attractor” solutions of inflationary theories for which the kinetic term has a pole in field space.  After lunch Jim Hartle discussed multiverses in quantum cosmology and Thomas Hertog described ongoing work with Stephen Hawking on obtaining a smooth exit from eternal inflation. Raphael Bousso argued that theorems in classical GR can be used to motivate conjectures about quantum field theory. He explained how one such conjecture can be proved for a class of quantum field theories. Finally, Renata Kallosh discussed amplitudes and finiteness of maximal supergravity, motivating models of inflation from fundamental theory. The day ended with a dinner at Trinity College at which Fay Dowker and Jim Hartle gave speeches recalling their time spent working with Stephen. 

Tuesday morning's talks were devoted to gravitational waves, following LIGO's sensational discovery of gravitational waves from merging black holes. Pablo Laguna discussed black hole kicks and possible observational signatures. Harald Pfeiffer describing different approaches to the modeling of gravitational waves from black hole binaries. Gabriela Gonz\'alez then gave an overview of LIGO's results so far. Bruce Allen discussed how Stephen and Gary Gibbons made serious attempts to build a gravitational wave detector in Cambridge. In the afternoon, Frans Pretorius argued that by combining the observations of multiple binary black hole mergers it might be possible to extract information about several quasinormal modes during the ``ringdown'' phase of the final black hole and thereby test the no-hair theorem. He also emphasized the challenge in modeling gravitational wave sources in alternative theories of gravity to further observational tests of general relativity. Mihalis Dafermos gave an overview of the cosmic censorship conjectures. Finally Ted Jacobson reviewed the subject of Hawking radiation from ``analog'' black holes and Jeff Steinhauer described his experimental work aimed at detecting analog Hawking radiation in a system involving a Bose-Einstein condensate.  

On Wednesday, Douglas Stanford discussed his work on traversable wormholes in the AdS/CFT correspondence. Gary Gibbons discussed the gravitational memory effect, and its relation to work he did with Stephen. Andy Strominger gave an overview of his work on infrared effects in QED and quantum gravity. Gary Horowitz argued that the AdS/CFT correspondence implies that it is impossible to pass through certain types of singularity in quantum gravity. He also discussed a new type of supersymmetric black hole solution, with topology outside the horizon, and explained why it presents a puzzle for microscopic calculations of black hole entropy. 

The conference ended with some closing remarks by Bob Wald. After praising Stephen's many contributions to gravitational physics, Bob explained in some detail why Stephen's paper ``Particle Creation by Black Holes'' is such a monumental achievement. However, he did identify one flaw in the paper: the word ``gauge'' is misspelled four times!

The conference talks can be viewed online at\\ 
\htmladdnormallink {\protect {\tt {http://www.ctc.cam.ac.uk/activities/stephen75}}}
{http://www.ctc.cam.ac.uk/activities/stephen75}

\vfill\eject

\section*{\centerline
{﻿Gravity: New perspectives from strings and higher dimensions}}
\addtocontents{toc}{\protect\medskip}
\addcontentsline{toc}{subsubsection}{
\it  Benasque workshop 2017, by Mukund Rangamani}
\parskip=3pt
\begin{center}
Mukund Rangamani, University of California, Davis
\htmladdnormallink{mukund-at-physics.ucdavis.edu}
{mailto:mukund@physics.ucdavis.edu}
\end{center}

The fifth edition of the biannual Benasque workshop entitled Gravity: New Perspectives from Strings and Higher Dimensions was held at the Pedro Pascual Centro de Ciencias between July 16 and 28, 2017.
The workshop was aimed at bringing together experts in general relativity and applications of gravitational physics to diverse physical systems as we have come to appreciate in the context of holography. 

The workshop, which now appears to be well established in the conference calendar, attracted a large number of participants, with expertise in a diverse set of fields, ranging from classical general relativity, numerical relativity, to holography and quantum information. The talks and discussions reflected this diversity to a large extent. To encourage free-form discussions and collaborations which have become the hallmark of the Benasque workshop, the program was kept relatively light, with 2 official talks in the morning. Over the course of the workshop, various participants organized discussions on topics of interest in the afternoon to further  foster collaborations.

The broad themes discussed during the workshop are described in some detail below:

\paragraph{String theory and black holes:} The workshop started off with a talk by Emil Martinec on the 
`String Theory of Supertubes' The idea was to examine the worldsheet theory in supertube backgrounds to examine whether stringy corrections are important in the black hole microstate geometries that have been constructed over the years. Finn Larsen, described his recent work on understanding  `Logarithmic Corrections to Black Hole Entropy'. By suitably embedding non-supersymmetric black holes (say Kerr-Neumann) in $\mathcal{N}=2$ supergravity, one was able to compute the 1-loop determinant for the fluctuations around the semi-classical saddle point, and furthermore argue that the answer which determines the logarithmic correction to black hole entropy is constrained by the embedding. We also heard from Manuela Kulaxizi about attempts to use conformal bootstrap techniques, to analyze the Regge limit of field theories and learn therefrom the constraints on the bulk dynamics (using holography). The basic idea is to understand the constraints from field theory for the dual gravitational system to be described by Einstein-Hilbert gravity. 

\paragraph{Applications of AdS/CFT:} There were several discussions on this topic, with Toby Wiseman
presenting his results on `On Universality of Holographic Results for (2 + 1)-Dimensional CFTs on Curved Spacetimes'
wherein he described the features that one may obtain geometrically (e.g. eigenvalues of the scalar Laplacian, or volumes of bulk (sub)-manifolds) which he then argued could be universal in CFTs. Sera  Cremonini described her work on constructing exotic phases of materials using holographic modeling in  her talk: `Intertwined Orders in Holography: Pair and Charge Density Waves'.

\paragraph{Entanglement, Complexity, and all that:} The field of holographic entanglement has evolved over the past few years in various directions. Not only are people investigating questions relating to emergence of geometry from entanglement, but there also has been some effort to import ideas of computational complexity into physics, and geometrizing them via AdS/CFT. Attempts in the latter direction were described by Adam Brown in his talk
`Complexity and Geometry', and we heard further on how to define complexity in quantum field theories (the definition in quantum mechanics is based on using a restricted set of quantum gates which act on a finite subset of constituents) of this in a discussion led by Michal Heller. In addition Dan Roberts talked about 
`Operator growth in quantum chaotic systems', which explored how simple operators grow to be complex under Heisenberg evolution. A related discussion on out-of-time order correlation functions in quantum systems and their potential applications was led by Felix Haehl and Mukund Rangamani.

The idea of entanglement providing the backbone for the emergence of geometry in the holographic context, implies as a corollary that the structure of entanglement also determines gravitational dynamics. Efforts to understand how this works were described in Felix Haehl's talk `Non-linear gravity from entanglement in CFTs'. In addition, Guifre Vidal spoke about tensor network models and lessons one can extract therefrom in his talk `Geometry without holography'. 
Aitor Lewkowycz gave a broad overview of recent attempts to understand construction of bulk geometric structures using modular evolution in his talk `Subregions and Quantum Gravity', and this was also followed up by a discussion led by him and Alex Belin.  Finally, we also learned about attempts to understand the Quantum null energy condition 
in dynamical settings in  Wilke Van der Schee's talk `Extremal surfaces, entanglement entropy and energy conditions from dynamical spacetimes'.

\paragraph{Classical Relativity}: The past few years have seen a great deal of progress (both theoretically and experimentally) in the realm of classical relativity. We have come to understand better various stability questions, developed new asymptotic expansion schemes that provide insight into gravitational dynamics, and are poised to make progress in getting a better handle on the strong gravity regime. Many of these topics were addressed during the workshop by several speakers. 

Benson Way described progress in understanding the stability of AdS spacetimes when perturbations carry rotational quantum numbers in his talk, `Instability of AdS with Angular Momenta', while Stephen Green gave a broad overview of the progress made in understanding the non-linear gravitational dynamics and the cause for AdS instabilities in  
his talk entitled `Nonlinear Dynamics in AdS'. In addition Hans Bantilan and Pau Figueras led a broad discussion on black hole instabilities in various systems, what is known about their onset, and the purported end-points. 

Valeri Frolov described recent progress in understanding the hidden symmetries in black hole spacetimes (Killing-Yano tensors, etc) in his talk `Black Holes, Hidden Symmetries, and Complete Integrability'. We also heard from Pavel Kovtun on a first principles treatment of `Relativistic magnetohydrodynamics'. On the analytic front, the large D expansion is proving to be valuable in a wide variety of circumstances, as Roberto Emparan reviewed in a discussion session. Amos Yarom applied this expansion scheme to understand the `The large D limit of turbulence'.

With the excitement of gravitational wave detection by LIGO, we closed out the program with two extremely topical talks:
Helvi Witek spoke about `Testing strong gravity in the gravitational wave era' while Shahar Hadar explained to us how  one may be able to discern signals from extremal black holes in his talk,  `Extreme Black Holes and Their Gravity Wave Signatures'.

Overall, the broad spectrum of talks and the time available for discussions was very much appreciated by the participants. Many of the participants were looking forward to the next iteration of the workshop which is scheduled to talk place in 2019 right after the Strings meeting in Brussels and the GR22 meeting in Valencia.

\section*{\centerline
{﻿Quantum Information in Quantum Gravity}}
\addtocontents{toc}{\protect\medskip}
\addcontentsline{toc}{subsubsection}{
\it  QIQG 3, by Matt Headrick and Rob Myers}
\parskip=3pt
\begin{center}
Matt Headrick (Brandeis University) and Rob Myers (Perimeter Institute)
\htmladdnormallink{mph-at-brandeis.esu}
{mailto:mph@brandeis.edu}
\htmladdnormallink{rmyers-at-perimeterinstitute.ca}
{mailto:rmyers@perimeterinstitute.ca}
\end{center}

The third conference in the series ``Quantum Information in Quantum Gravity'' was held August 14-18 on the beautiful campus of the University of British Columbia in Vancouver. During the three years since the first conference in the series, also held at UBC in 2014, concepts from quantum information theory---such as entanglement, quantum error correction, quantum complexity, and tensor networks---have become ever more mainstream tools in current research in quantum gravity. This conference brought together around 80 researchers, including those with backgrounds both in quantum information theory and in various aspects of quantum gravity such as string theory, as well as young people who have grown up fluent in both languages.

2017 marks 20 years since the birth of the AdS/CFT correspondence, which has revolutionized our ideas about both quantum gravity and strongly-coupled quantum field theories, and it certainly played a central role in the work presented at the conference. Concepts from quantum information theory have been instrumental in our ongoing attempts to ``decode'' this mysterious duality. A key development was the discovery of the holographic entanglement entropy (EE) formula by Ryu and Takayanagi in 2006, which led to the concept of entanglement-wedge reconstruction, and from there to the view of holography as a quantum error-correcting code. Advances along these lines were discussed in many of the talks. Matthew Headrick explained how the ``bit-thread'' view of holographic EE could be made fully covariant. Jennifer Lin explained how Ryu-Takayanagi EE could be naturally viewed as arising from edge modes. Antony Speranza tackled the tricky problem of quantifying those edge modes in diffeomorphism-invariant theories. Charles Rabideau and Felix Haehl described progress in the program of deriving the Einstein equation from the Ryu-Takayanagi formula together with properties of EEs. Hirosi Ooguri used the notion of entanglement-wedge reconstruction to argue for several ``swampland'' conjectures, which constrain possible UV-completable quantum theories of gravity, such as the absence of global symmetries and the weak gravity conjecture. Netta Engelhardt proposed an interpretation for the area of black-hole apparent horizons, namely as the maximal entropy of any black hole with the same geometry outside the apparent horizon. Aitor Lewkowycz employed the concept of ``modular flow'' to make more precise the notion of entanglement-wedge reconstruction. Tom Hartman reviewed recent developments using the Ryu-Takayanagi formula to relate fundamental properties of the boundary CFT, including causality, the average null energy condition, and the quantum null energy condition. And Ning Bao explained how to use the Ryu-Takayanagi formula to evaluate the Holevo information in thermal states, and thereby quantify how distinguishable black hole microstates are if one only has access to a portion of the boundary. Trying to pin down the precise meaning of the phrase ``entanglement-wedge reconstruction'' (or ``subregion/subregion duality'') was the topic of a lively discussion led by Veronika Hubeny.

Another popular direction of AdS/CFT research, also originally inspired by the Ryu-Takayanagi formula, is based on the idea that the bulk spacetime can in some sense be viewed as a tensor network that prepares the ground state (or other state) of the CFT. Tensor networks in turn inspired the conjecture that there is a relation between the volume of a spatial slice of the bulk and the quantum complexity of the CFT state (measured by the number of gates required to produce that state). Both tensor networks and complexity (and the relation between them) were major themes of the conference. Xiao-Liang Qi talked about some recent efforts to understand the nature of causal influences in the context of tensor networks. In part, his work was motivated by rival claims that MERA networks gave a discrete description of a spatial slice of AdS space and that they describe a de Sitter geometry. His message was that the question of how an operator insertion ``influences'' a network depends on not just the position of where the influence is measured but also on the locality or nonlocality of the measurement. Adam Brown described the ``complexity = action'' conjecture and reviewed the evidence for it. This is a refinement of the ``complexity = volume'' conjecture mentioned above, which relates the gravitational action of the Wheeler-de Witt patch of a boundary time-slice to the complexity of the state of the CFT on that slice. Tadashi Takayanagi proposed that the bulk spacetime in AdS/CFT arises from optimizing a (continuous) tensor network preparing the state of the CFT to minimize its complexity. Bartek Czech described closely-related ideas of deriving the equations of motion for three-dimensional gravity from varying complexity. This work draws inspiration from tensor networks but aims to provide a clearer definition of complexity in conformal field theories.

AdS/CFT has often been used to try to understand the paradoxes around black holes. However, as emphasized by Jonathan Oppenheim, so far it has served more to sharpen than to resolve those paradoxes. Indeed, using the information-theoretic notion of ensemble ambiguity, he argued that there is an inconsistency in the conventional wisdom regarding eternal black holes in AdS representing thermofield-double states in the CFT.

A recent development in AdS/CFT that has attracted considerable attention involves the so-called Sachdev-Ye-Kitaev (SYK) model, a solvable quantum-mechanical model that is dual to a theory of two-dimensional AdS gravity. We had several talks related to the SYK model. Dan Roberts described how operators ``grow'' in the model under Heisenberg evolution, and how this growth explains the scrambling behavior that the model displays. Douglas Stanford's talk had two parts. In the first, he discussed the appearance of firewalls and the state dependence of operators in black hole states and in the second, he described some higher dimensional versions of the SYK model. And Javier Magan described his investigation into the properties of EE in large-N theories, with particular attention to theories with a dual gravity description, such as the SYK model and large-N vector and gauge theories.

One of the more intriguing questions of the conference came during a discussion session chaired by John Preskill, during which he pointed out that, thanks to rapid advances in quantum computational hardware, it was likely that within a few years it will be possible to simulate the strongly-coupled quantum field theories appearing in AdS/CFT---in other words, to simulate a full theory of quantum gravity! The question of what we would hope to learn from such simulations was left as a homework problem for the conference participants.

As illustrated by the above discussion, the AdS/CFT correspondence, in its various forms, continues to dominate fundamental work on quantum gravity, especially in its connections to quantum information theory. Nonetheless, several talks at the conference explored other directions. Djordje Radicevic described a novel type of renormalization-group analysis for quantum-mechanical systems that do not necessarily have a spatial structure, and, similar to the SYK model, speculated that it offered a promising way to generate quantum gravitational theories from simple spin systems. Sean Weinberg described a generalization of the Ryu-Takayanagi formula to arbitrary (not necessarily asymptotically AdS) spacetimes, involving so-called holographic screens, and discussed an ongoing effort to understand the quantum information-theoretic structure it implies. Beni Yoshida followed up on work by Hayden and Preskill showing that the fast-scrambling property of black holes implied that information dropped into a black hole came back out immediately in the Hawking radiation. Yoshida showed that it is in some sense ``easy'' to decode the information from the radiation. The protocol to do so is in fact closely related to the procedure for making a traversable wormhole by coupling the two sides of an eternal black hole recently discovered by Gao, Jafferis, and Wall.

The very last session of the conference was devoted to a discussion led by Mark Van Raamsdonk on ``Quantum Gravity in General Spacetimes''. The AdS/CFT correspondence was discovered 20 years ago and in this time, we have made great strides in understanding the nature of quantum gravity in the presence of a negative cosmological constant, ie, for asymptotically anti-de Sitter spacetimes. However, it is also natural to investigate how to work with quantum gravity in more general situations, ie asymptotically flat or de Sitter spaces or more general cosmological spacetimes. In particular, of course, our present universe appears to be dominated by a positive cosmological constant and so we expect to be in a de Sitter phase. We began by discussing the challenges that these more general situations present: For example, one might ask to what extent holography is a useful tool to apply in general spacetimes. More generally, what is the correct mathematical structure that underlies quantum gravity, eg, should we still be considering states in a Hilbert space? While we have learned a great deal about the emergence of space from the AdS/CFT correspondence, what about the emergence of time or causal structures? We also discussed useful hints and ideas coming from our recent progress with AdS/CFT which might be applied in more general situations. One suggestion was that the application of tensor networks and quantum error correcting codes seem to hold more general lessons connecting the structure of spacetime and the structure of the corresponding quantum state. Unfortunately the list of challenges seemed to be much longer than the list of new ideas on how to proceed, however, Mark closed by encouraging us all, but especially the younger people, to think about these hard problems from time to time. Even if we make a small amount of progress, it is certainly a worthwhile direction to move. 

The workshop gave us a stimulating week of talks and discussions, which reflected the ongoing excitement in the field. For the interested reader, let us add that the fourth edition of the conference series will be held in Florence, Italy at the Galileo Galilei Institute, June 11-15, 2018. 

\vfill\eject
\section*{\centerline
{Remembering Cecile DeWitt-Morette}}
\addtocontents{toc}{\protect\medskip}
\addtocontents{toc}{\bf Obituary:}
\addcontentsline{toc}{subsubsection}{
\it  Remembering Cecile DeWitt-Morette, by Pierre Cartier}
\parskip=3pt
\begin{center}
Pierre Cartier, IHES Paris 
\htmladdnormallink{cartier-at-ihes.fr}
{mailto:cartier@ihes.fr}
\end{center}

I. The eventful life of C\'ecile (1922-2017)

II. My collaboration with C\'ecile

\vglue 1cm

\noindent {\bf I.	The eventful life of C\'ecile (1922-2017)}

\medskip

C\'ecile was born on December 21, 1922, in an apartment of the Ecole des Mines\footnote{If I remember correctly, her grandfather was the general secretary of the school.} facing the beautiful Luxembourg Garden in Paris. Her father was an important industrialist who died early. Her mother remarried with the principal collaborator of her husband.  C\'ecile spent the first years of her life in Mondeville, in Normandy, followed by a year in Paris, where she was enrolled in second grade with her future friend and co-author Yvonne Choquet-Bruhat.  After the year in Paris, her family moved to Caen, also in Normandy.  She lived in Caen through most of World War II, and happened to be safe in Paris taking a physics exam on D-Day, but a bomb fell on her home in Caen and killed her grandmother, her mother, her beloved 16-year old sister, and the cook. We have to recall that Caen was heavily bombed in June 1944 by the Allied forces. Moreover, after the complete invasion of France, in the fall 1942, by German and Italian troops, France was divided into 8 different zones and an ``Ausweis'' ($=$ passport) was required for interzone travel. C\'ecile had been given permission to travel between Caen and Paris to attend de Broglie's lectures. That explains how she escaped death.

After the war, she began working  on a thesis under L.~de~Broglie. Soon after, she was hired by Fr\'ed\'eric Joliot(-Curie). Joliot, who, after surviving with some difficulty during the Nazi invasion of France, had become the head of the France nuclear program, as well as a fanatical communist militant, was too busy to prepare his lectures at Coll\`ege de France, and delegated the preparation to the rather inexperienced C\'ecile. Then she visited Great Britain and Ireland with the mission of interviewing important physicists about their work during the war. After a memorable encounter with Dirac, more talkative than customary, she visited the Institute for Advanced Study in Dublin. There she met E. Schr\"odinger, who had escaped from nazi Austria with his ``two wives" and was welcomed by the Cardinal! There was also W. Heitler, who suggested to C\'ecile to work on meson physics, especially the distinction between $\pi$ and $\mu$ mesons. That was the subject of her thesis\footnote{With title: ``Sur la production des m\'esons dans les chocs entre nucl\'eons''.}, defended in front of de~Broglie, after her return in Paris.

While in Copenhagen, C\'ecile received a telegram from Robert Oppenheimer, which read, ``On the recommendation of Bohr and Heitler, I am glad to offer you membership in the school of mathematics Institute for Advanced Study (IAS) for the academic year 1948-1949 with a stipend of \$3500.00.''  Although C\'ecile had no idea where Princeton was, she accepted the invitation and there began her lifelong friendship with Freeman Dyson, and a more complicated relationship with Richard Feynman\footnote{His relationship with women was always complicated!}. It was at the IAS, the following year, that she met her future husband, Bryce Seligman DeWitt, a ``Schwinger boy'' and veteran of the US Navy, in which he had been trained to be a naval aviator during the war.  C\'ecile initially rejected Bryce's marriage proposal and accepted only after she came up with the idea, literally overnight, of creating the Les Houches summer school as her penance for marrying a foreigner and as a way to contribute something back to France.

At the time, in the fifties, experimental physics (and chemistry) was still strong in France, mainly as part of the Curie heritage. Nuclear physics, magnetism (remember the Weiss' domains\footnote{Henri Cartan was the son in law of Weiss!}), solid state physics (created by Pierre Aigrain and Alfred Kastler and developing later into ultralow temperatures by Claude Cohen-Tannoudji and his school). On the other hand, theoretical physics was in a poor state. The school of de~Broglie, filled with incompetent and sometimes dishonest people, went into a deadlock by refusing the ``Copenhagen interpretation'' and developing tools like ``the double solution'' leading nowhere. Louis de Broglie considered that quantum mechanics was like a luxurious jewel, to be reserved for the happy few. As a result, we had to wait until the late sixties for the creation of regular graduate (and then undergraduate) courses about quantum mechanics. In the fifties, neither
Landau and Lifschitz' textbook nor the Feynman lectures were available, and I learned quantum mechanics from the giant books of Hermann Weyl and P.A.M. Dirac.

Gravitation and General Relativity were not in a better situation. The founding fathers in France, namely Georges Darmois, Marie-Antoinette Tonnelat, and Andr\'e Lichnerowicz were skilled mathematicians, who developed it like a purely mathematical theory. Yvonne Choquet-Bruhat came later with more interest in physics. I remember discussing around 1970 with Lichnerowicz the first experiments by Weber. He was not at all interested despite his own mathematical contributions to the existence of gravitational waves. It was the team of theoretical physicists at Meudon (Brandon Carter, Thibault Damour) who transformed, within the French community, General Relativity into a physical discipline\footnote{It took fifty more years to realize the dream of observing these gravitational waves!}.

To fill this vacuum, C\'ecile came up with a very original and bold idea, to create a summer school in physics, where, in the inspiring place of Les Houches in the Alps, students and professors would share life and science for two months. This was started in 1951 and brought lecturers like Pauli or Feynman. The place has been a landmark in theoretical physics for 70 years.

For me, it was a ``missed opportunity'', which influenced my own career. The first sessions were attended by some of my classmates, like Pierre-Gilles de Gennes and Bernard Malgrange. I was then 22 and full of youthful dreams, ranging from theology (I had just given up the idea of becoming a minister in the French Calvinist Church), to unorthodox Marxist philosophy (following my professor of philosophy Louis Althusser) and to radioastronomy (then a new science). But in between, I had been seduced by Bourbaki, and put all of my energy in their ambitious program of reconstructing mathematics from scratch.

After Princeton and the marriage of C\'ecile and Bryce, came the period of Chapel Hill. For 16 years she was unable to get a position there, because of a rule forbidding a married couple to be hired in the same department. She was the mother of four daughters, as efficient as a mother as in any of her enterprises. One of her daughters has a mental illness. She was a big concern for C\'ecile, but when you are C\'ecile, the answer is to become actively engaged in mental health endeavors, establishing a local mental health support organization, and organizing fundraising events such as one honoring a prominent caregiver, Alicia Nash, who cared for her husband John Nash and their son through many years of their respective illnesses.

Then, in 1972, came the fortunate move from Chapel Hill to Austin. Austin is a very pleasant campus and downtown\footnote{Not everyone realizes that Austin is a dual city, the Mexican one on the eastside, and the ``WASP'' one on the westside, separated by the physical barrier of Interstate 35. Now that the airport has moved to the East side, visitors have more difficulties ignoring this geographical feature. Also, in the ``Cinco de Mayo'' (5 May, Mexican national festival), the town belongs to the Mexicans. Notice also that, in an academic building after 8 p.m., if you need a key, better ask about it in Spanish!}. On both gown and town you have a wealth of theatrical and musical events. Besides the ``President Johnson Library'' and the football field, you have a big auditorium seating 3000 persons, where you can watch the Covent Garden, three nights in a row, for \$20 a night.

Despite Texas' climate, the town is green, with many parks, an artificial lake and a ``natural'' pool in a canyon, where you swim in the midst of turtles (and perhaps snakes!). To sum up, Austin, both campus and town are liberal within an harshly conservative Texan environment. A standard saying is: ``If Texas secedes\footnote{I doubt that after the civil war, Texas would be permitted to secede, but there is a bylaw allowing Texas to split into five pieces, each one sending two senators to Washington, D.C. A similar bylaw applies to California.} some day from the United States, the day before Austin will secede from Texas''.

C\'ecile was at first appointed to the Department of Astronomy. The dominant figure there was G\'erard de Vaucouleurs\footnote{The Vaucouleurs had a very gifted technician, who decided once to run with his daughter a (very successful) business of French bakeries. Before leaving, he had to swear to visit Austin for a week every year: he was the only one able to unlock some secret tools!}, a French \'emigr\'e, who together with his wife maintained and developed a catalog of galaxies. Now that we estimate the number of galaxies to be of the order of $10^{11}$ (each galaxy with $10^{10}$ stars in the mean), I don't know the purpose and meaning of such a catalogue today. Many observations were done at the MacDonald Observatory, and people commuted between Austin and MacDonald (a roundtrip drive of 1500 km in the Texas desert). Nowadays, only people in charge of maintaining the instruments visit MacDonald, and astronomers watch the sky through their computer screen.

After some years -in 1983- she joined the Physics Department. This was a strong department, chaired at the time by Tom Griffy.  Yuval Ne'eman and Ilya Prigogine were part-time faculty members. Prigogine had received the Nobel Prize in 1977. Texas physics Professor Steven Weinberg, also a Nobel laureate, is a superb scientist, with a list of beautiful books (including one in cosmology), but a strong character who claims to know everything\footnote{It took me time and flexibility to be respected by him. I was helped in this endeavor by a visit of Ludwig Faddeev, who treated me like an old friend!}. Despite political differences\footnote{Ne'eman had many political and military responsibilities in Israel, and is the father of Israel's nuclear weapons program. In my opinion he would have deserved to be a Nobel laureate, but perhaps the Nobel committee was afraid of political controversies.}, I became a friend of Ne'eman and we shared students and seminars. On the other hand, Prigogine was disappointing. When he announced a seminar talk, it was eventually delegated to one of his assistants. Nonequilibrium physics is a fascinating subject, but he used the wrong tools for that\footnote{I once advised a graduate student in his group to go elsewhere after being granted his Ph.D.}.

A feature of the Physics Department at that time was its strong research centers. One was the Center for Relativity, which had been created by Alfred Schild. Bryce DeWitt was the Director in 1983. C\'ecile was a member of that Center. There was also the Center for Theoretical Physics, created by John Archibald Wheeler, after his retirement from Princeton University. And Steven Weinberg headed a third center, the Theory Center.  At this point, Wheeler was interested  in Quantum Gravity, as was Bryce DeWitt, who had been interested in Quantum Gravity from the beginning.  There was another character in this center, whom we called the ``captain'' because he was later on associated with joint experimental programs of the Navy and NASA (first versions of the GPS system!).

Both C\'ecile and Bryce were adventurous characters. Before telling their adventure in Mauritania to watch a solar eclipse, I have to insert a comment on the experimental basis of Einstein's gravitation theory. Three tests of General Relativity have been historically important:

a) {\it The residual motion of Mercury's perihelion} ($43''$ of arc per century!). Despite many efforts, and many proposals (a new planet, Vulcan, between the Sun and Mercury, an oblate Sun, with quadrupolar effect, modification of Newton's law) no astronomical explanation was successful, while Einstein's calculations in 1915 fitted the data very well.

b) {\it The redshift of spectral lines}. Stellar data were quite inconclusive, and only through the use of the Mossbauer effect on earth the experiment of Pound and Rebka in 1959 confirmed this redshift, now a central element in astrophysics!

c) {\it The bending of light rays around the Sun}. Here there is a crucial factor of 2. According to a calculation going back to Laplace, and assuming that a light ray corresponds to the motion of a particle (photon?), the gravitational influence of the Sun results in a Keplerian hyperbola, independent of the mass of the photon by Galileo's universality of free fall. The deviation, that is the angle between the two asymptotic directions, is $2\alpha$ radians, in the first order in $\alpha = \frac{GM_{\odot}}{bc^2}$ ($G$ gravitation constant, $c$ speed of light {\it in vacuo}, $b$ impact parameter, $M_{\odot}$ mass of the Sun). Numerically $\alpha \sim 2 \times 10^{-6}$ for a ray grazing the Sun.

In 1907, Einstein tried for the first time to account for the influence of gravitation on light. He assumed that at a place ${\bf x}$ where the gravitation potential is $\Phi$, the speed of light is $\sqrt{c^2 + 2 \Phi}$ (a consequence of the conservation of energy for a particle!). To include relativity effects, we have to modify the Minkowski metric $\eta = c^2 \, dt^2 - d{\bf x}^2$ by replacing $c^2$ by $c^2 + 2\Phi$. If we do not change the spatial metric $d{\bf x}^2$, we get in first approximation in $\alpha$ the same result, $2\alpha$ radians, as the one given by the above-quoted Newtonian reasoning. In polar coordinates, we have $d{\bf x}^2 = dr^2 + r^2 (d\theta^2 + \sin^2 \theta \cdot d \varphi^2)$. But taking in full account the field equations for gravitation (discovered by Einstein in November 1915!), we have to replace $dr^2$ by $dr^2 / \left( 1+2 \frac{\Phi}{c^2} \right)$ with the gravitational potential $\Phi = - GM_{\odot} / r$ (Schwarzschild). To calculate the deviation to first order in $\alpha$, it is enough to use an approximate solution discovered by Einstein in 1915. If we take this into account, the deviation becomes $2 \times 2 \alpha$ radians (equal to $1.75''$ of arc). Here is the factor of $2$. Eddington in 1919 claimed to have observed this. 

Let us go back to Bryce and C\'ecile! They decided to improve on previous measurements of light deflection by taking advantage of the 1973 solar eclipse, visible in Mauritania. Their equipment was transported by the U.S. National Center for Atmospheric Research, which arranged a large ($\sim$~50 person) eclipse expedition to a desert oasis, Chinguetti, Mauritania, including six scientists carrying out the deflection of light measurement (among them C\'ecile and Bryce.)\footnote{Let me add that Bryce was a well-seasoned pilot from his war experience. In the beginning of space exploration, he volunteered and went through the physical tests, only to be turned down because he was over forty.}
Their observations confirmed Einstein's prediction to ~10\%. 

Summarizing, there was  little solid observational basis for General Relativity before 1960.
Things have changed: within the solar system the gravitational field is of order $\alpha$ at most (weak field) but the post Newtonian approximation program, with data fitting numerous space observations, gives an excellent agreement. For strong fields, binary pulsars and more recently, gravitational waves, give unambiguous, accurate confirmations.

Before describing our collaboration, let me give a few examples of the managerial ability of C\'ecile.

Once in Les Houches, during one of my attendances there, there was no one to serve breakfast on a Sunday morning: the two local aids in the kitchen had been out dancing for the whole night. Ten minutes later, everything was in order with C\'ecile herself preparing coffee and toast!

Much later, there were discussions about recruiting her for the board of trustees of IHES, and some people objected about her age, being afraid that she would sleep during the meetings. I told everyone that no one can sleep in a meeting where she is present! Indeed, she usually came a few days in advance, meeting everyone on the staff to know everything by herself. One of her last scientific works was a publication of Bryce's papers\footnote{{\it The Pursuit of Quantum Gravity. Memoirs of Bryce DeWitt}, Springer, 2000.}.

\bigskip 

\noindent {\bf II.	My collaboration with C\'ecile}

\medskip

It lasted more than 30 years and ended with the publication of a book\footnote{P. Cartier and C. DeWitt-Morette, {\it Functional integration (Action and symmetries)}, Cambridge Univ. Press, 2006.}. It started in two distinct ways. When C\'ecile visited Princeton in the 1950's the latest development in physics was {\it Quantum Electrodynamics}. What was needed was a mathematical model to account for the newly observed Lamb shift, at a deeper level account for the radiative corrections to atomic spectra. In classical terms, the reaction on an electron of mass $m$ of the electromagnetic field emitted by this electron is infinite (for a pointlike electron). In quantum theory, the effect is still infinite, but with a logarithmic (and not linear) divergence. Bethe proposed a {\it cutoff}: the frequencies higher than a bound $\Lambda = \frac{mc^2}{h}$ should be ignored. The challenge was to make it compatible with (special) relativity. Using a very ad hoc proposal (``path integrals'' or ``sum over histories''), R. Feynman gave a recipe using his well-known diagrams. For instance, for one incoming and one outgoing electron, intermediate stages can be pictured by diagrams like
$$
\includegraphics[width=75mm]{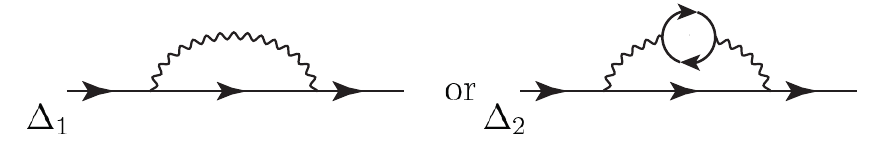}
$$
(solid lines for electrons, wavy lines for photons). If the incoming (and outgoing) 4-momentum is ${\bf p}$, then each diagram $\Delta$ contributes an amplitude $A_{\Delta} ({\bf p})$ and one estimates probabilities by $\left\vert \underset{\Delta}{\sum} \ A_{\Delta} ( {\bf p}) \right\vert^2$. Feynman gave explicit formulas to calculate $A_{\Delta} ({\bf p})$ by means of integrals. The complexity of these integrals increases with the number of {\it loops} in the diagrams (1 for $\Delta_1$, 2 for $\Delta_2$, etc$\ldots$). The result can be expanded into a power series in the coupling constant $\alpha_e = \frac{e^2}{4 \pi \varepsilon_0 \hbar c}$. Each such integral diverges logarithmically, but putting a Bethe-type cutoff $\Lambda$ makes them convergent, and the recipes of renormalization tell us how to go to the limit $\Lambda \to \infty$.

That was an immediate success, most notably by the calculation of the Lamb shift and anomalous magnetic moment of the electron. The intuitive character of the diagrams made them very popular, but Feynman's justification by path integrals seemed ad hoc, and was ignored by most people. Soon after, F. Dyson rediscovered under the name of ``time-ordered exponential series'' a familiar tool in ordinary differential equations (Lappo-Danilevsky), and gave an {\it ab initio} derivation of Feynman's formulas.

Some years later, Feynman gave a basic course about path integrals, which eventually was written up by his student Hibbs and published under their two names. This was not a success, and the book remained very unconvincing. Why? I don't know. There were numerous attempts by both physicists and mathematicians, in the West as well as in Soviet Union, to formulate correctly the path integrals. But before describing the proposal made by C\'ecile and myself in our book, let me describe the other development, running in parallel: {\it history of measure theory and Lebesgue integral.}\footnote{A readable account is given by Bourbaki (see {\it ``El\'ements d'histoire des math\'ematiques''}, Hermann, Paris, 1960, $2^{\rm nd}$ \'edition, Springer Verlag, 2007).}

Lebesgue integral was invented around 1900, and underwent many developments until 1940. A notable event was the proof by Markoff in 1938 of a vast generalization of previous results of J. Hadamard and F. Riesz. This result establishes a bijection, between regular Borel measures\footnote{A Borel measure on a compact space $X$ is a nonnegative $\sigma$-additive function $\mu$ on the $\sigma$-algebra of Borel subsets of $X$. Regularity means that for any measurable subset $E$ of $X$ and any $\varepsilon > 0$, there exist two subsets, $K$ compact, $U$ open such that $K \subset E \subset U$ and $\mu (U) - \mu (K) \leq \varepsilon$. Regularity is automatic if there is a countable basis of open sets in $X$.} $\mu$ on a compact space $X$, and the linear forms $\overline\mu$ on the Banach space $C^0 (X ; {\mathbb R})$ of continuous real-valued functions on $X$, which satisfy $\overline\mu ( f ) \geq 0$ for $f \geq 0$. The correspondence is given by $\overline\mu ( f ) = \int_X f \cdot d\mu$. As a bonus, the Lebesgue space $L^p (X,\mu)$ is the completion of the space $C^0 (X ; {\mathbb R})$ under the norm $N_p(f) = \left( \int_X \vert f \vert^p d\mu \right)^{1/p}$. This result is readily extended to a locally compact space, and A. Weil saw this duality approach as the right foundation for a general integration theory. His views were supported within Bourbaki by L. Schwartz, who liked the idea of duality and used it in the creation of the theory of distributions. R. Godement put all of his energy to pursue the project which ended by the publication of Bourbaki in several volumes around 1950.

The method worked well for {\it locally compact} spaces. But Bourbaki was obsessed by the idea of removing any countability assumption from topology and analysis -- at the cost of rather akward definitions, like the one of measurable sets or functions (also the various definitions of the exterior measure of a set!). The limitation to locally compact spaces didn't look serious for geometrical and even number-theoretical applications.

As long as probability theory focused on finite (or even countable) families of random variables, one needed only integration over finite-dimensional manifolds (hence locally compact spaces) but the rapid development of random processus required new tools, and Bourbaki decided to complete his treatise on integration. This was done in the 1960's. Officially, every text in the Bourbaki treatise is anonymous, and it is really a team work. Today, I can say that the impetus came from L. Schwartz, P.A. Meyer and myself. Schwartz had as father-in-law, Paul L\'evy, a famous probabilist, P.A. Meyer was then a star among French (and world) probabilists, and myself was motivated by my interest in probability and mathematical physics. After many strenuous efforts, it was generally accepted that the right category of spaces were the {\it Polish spaces} (separable, complete, metric) or some slight generalizations (Souslin or Lusin spaces). Borel measures were automatically regular and the Soviet School (Gelfand, Prokhorov, Minlos,~$\ldots$) had established the existence of infinite products of measures, or more generally projective limits. There a compactness criterion, given by Prokhorov, gives the final answer.

The most innovative section in the new chapter of Bourbaki deals with {\it promeasures}, a linear version of the general method sometimes called cylindrical measures\footnote{This corresponds to the habit in probability theory, of describing the law of a process $(X(t))_{t \in T}$ by giving the {\it marginals}, that is the law of random vectors $(X(t_1) \ldots X(t_n)) \in {\mathbb R}^n$ extracted from the random process.}. Following S. Bochner (1955) and generalizing a method invented by P. L\'evy in the 1920's one introduces the {\it Fourier transform} of a promeasure in the linear space $X$, a function on the dual space $X'$ of $X$. By Minlos theorem, if the space $X$ is {\it nuclear} in the sense of Grothendieck's thesis, the continuity of the Fourier transform on $X'$ is what is needed to have a true measure on $X$, instead of the approximating promeasure.

Here the two trends came together. It started in an exchange between J.~Dieudonn\'e and C\'ecile in 1971. She was mentioning to him that in physics one needs oscillating integrals like $\int_{-\infty}^{+\infty} e^{ix^2} dx$ instead of convergent integrals like $\int_{-\infty}^{+\infty} e^{-x^2} dx$. Dieudonn\'e replied that Fourier transforms exist for distributions and take care of some of the oscillating integrals. His suggestion was to replace {\it pro-measures} useful in probability theory by {\it pro-distributions} and to ask Cartier about that. A week later, I met C\'ecile at a party at IHES, and so started our long-term collaboration and friendship. So, I could repair the ``missed opportunity''. At last my investment in writing the Bourbaki treatise could bring fruits for our collaboration.

Our scientific marriage lasted more than 30 years. C\'ecile had already published two books\footnote{Y. Choquet-Bruhat and C. DeWitt-Morette, {\it Analysis, Manifolds and Physics}, Part I: {\it Basics} and part II: {\it 92 applications}, North Holland, 1982 and 1989.} in collaboration with Yvonne Choquet-Bruhat and the first idea was a book by her and her husband, together with mathematical appendices by myself. This could not be done for scientific and personal reasons, and soon we agreed on writing together a book\footnote{For a technical description of this book, see the companion note ``A tutorial in Feynman path integrals''.}, while Bryce DeWitt published his own version\footnote{B. DeWitt, {\it The global approach to Quantum Field Theory}, Oxford Univ. Press, 2004.}. It required me to visit Austin every year, visits I enjoyed very much. I was the guest of both the Physics and Mathematics departments. On both sides, I had good connections, most notably John Tate on the Math. department. I shared graduate students with C\'ecile. I particularly remember a young Chinese woman by the name of Xiao-Rong Wu, who needed less than a year to transform from a shy Chinese woman into a standard American student, marrying an American. Also, a German by the name of Saemann (whom I met recently in Edinburg) and Markus Berg a Swede who came back to his country because he was not a member of the ``string fashion'' in the USA; he offered a wonderful wedding party in Austin. C\'ecile and myself organized many scientific events in Les Houches, Cargese, Institut Henri Poincar\'e, Strasbourg, etc. $\ldots$ My wife and I visited Texas, Louisiana, New-Mexico in memorable travels and as mentioned already, I enjoyed life in Austin: bicycle, swimming, concerts, theater plays, the bats under the bridge on the lake, $\ldots$ Too much to give here details. For all that, including the rewarding enterprise of writing a book:

\vglue 1cm

\centerline {MERCI C\'ECILE}
\bigskip
{\it Editor's note: \\
Friends, former students, and family of Professor C\'ecile DeWitt-Morette and Professor Bryce DeWitt have joined together to establish the C\'ecile DeWitt-Morette and Bryce DeWitt Endowed Graduate Fellowship in Physics at The University of Texas at Austin.  See}
\htmladdnormallink
{\protect {\tt{https://cns.utexas.edu/cecile-dewitt-morette-and-bryce-dewitt-\\endowed-graduate-fellowship-in-physics}}}
{https://cns.utexas.edu/cecile-dewitt-morette-and-bryce-dewitt-endowed-graduate-fellowship-in-physics}
\end{document}